\documentclass[floatfix,
reprint,
superscriptaddress,
amsmath,amssymb,
aps,
prl,
]{revtex4-2}

\usepackage{graphicx}
\usepackage[dvipsnames]{xcolor}
\usepackage{pdfpages}
\usepackage{amssymb}
\usepackage{amsmath}
\usepackage{amsbsy}
\usepackage{color}
\usepackage[normalem]{ulem}
\usepackage{epstopdf}
\usepackage{bm}
\usepackage{xurl}
\usepackage{textcomp}

\usepackage{float}

\newcommand{\dm}[1]{\textcolor{Black}{#1}}

\newcommand{\ldna}{$\lambda$DNA~}

\graphicspath{{figures/}}

\usepackage{xr}
\makeatletter
\AtBeginDocument{\let\LS@rot\@undefined}
\newcommand*{\addFileDependency}[1]{
  \typeout{(#1)}
  \@addtofilelist{#1}
  \IfFileExists{#1}{}{\typeout{No file #1.}}
}
\makeatother
\newcommand*{\myexternaldocument}[1]{%
    \externaldocument[SM-]{#1}%
    \addFileDependency{#1.tex}%
    \addFileDependency{#1.aux}%
}
\myexternaldocument{si}

\begin{document}
\title{NANOG assembles into self-limiting aging micelles \\ that drive a sol-gel transition and modulate DNA dynamics}

\author{Amandine Hong-Minh}
\thanks{Joint first author}
\affiliation{Centre for Regenerative Medicine, Institute for Regeneration and Repair, 5 Little France Drive, Edinburgh EH16 4UU, Scotland}
\affiliation{School of Physics and Astronomy, University of Edinburgh, Edinburgh Eh9 3FD, Scotland}

\author{Yair Augusto Guti\'{e}rrez Fosado}
\thanks{Joint first author}
\affiliation{School of Physics and Astronomy, University of Edinburgh, Edinburgh Eh9 3FD, Scotland}

\author{Abbie Guild}
\affiliation{School of Molecular Biosciences, University of Glasgow, University Avenue, Glasgow G12 8QQ, Scotland}
\affiliation{Centre for Regenerative Medicine, Institute for Regeneration and Repair, 5 Little France Drive, Edinburgh EH16 4UU, Scotland}
\affiliation{Institute for Stem Cell Research, School of Biological Sciences, University of Edinburgh, 5 Little France Drive, Edinburgh EH16 4UU, Scotland}

\author{Nicholas Mullin}
\affiliation{Centre for Regenerative Medicine, Institute for Regeneration and Repair, 5 Little France Drive, Edinburgh EH16 4UU, Scotland}
\affiliation{Institute for Stem Cell Research, School of Biological Sciences, University of Edinburgh, 5 Little France Drive, Edinburgh EH16 4UU, Scotland}

\author{Laura Spagnolo}
\affiliation{School of Molecular Biosciences, University of Glasgow, University Avenue, Glasgow G12 8QQ, Scotland}

\author{Ian Chambers}
\thanks{corresponding author, ichambers@ed.ac.uk}
\affiliation{Centre for Regenerative Medicine, Institute for Regeneration and Repair, 5 Little France Drive, Edinburgh EH16 4UU, Scotland}
\affiliation{Institute for Stem Cell Research, School of Biological Sciences, University of Edinburgh, 5 Little France Drive, Edinburgh EH16 4UU, Scotland}

\author{Davide Michieletto}
\thanks{corresponding author, davide.michieletto@ed.ac.uk}
\affiliation{School of Physics and Astronomy, University of Edinburgh, Edinburgh Eh9 3FD, Scotland}
\affiliation{MRC Human Genetics Unit, Institute of Genetics and Cancer, University of Edinburgh, Scotland}
\affiliation{International Institute for Sustainability with Knotted Chiral Meta Matter (WPI-SKCM$^2$), Hiroshima University, Higashi-Hiroshima, Hiroshima 739-8526, Japan}

\begin{abstract}
\textbf{Proteins and nucleic acids form non-Newtonian liquids with complex rheological properties that contribute to their function \textit{in vivo}. Here we investigate the rheology of the transcription factor NANOG, a key protein to maintain embryonic stem cell pluripotency. \dm{We find that at high concentrations, NANOG forms macroscopic aging gels that are dependent on its intrinsically disordered domain. By combining molecular dynamics simulations, mass photometry and Cryo-EM, we also discover that -- in contrast with unbounded condensates formed by other intrinsically disordered proteins -- NANOG forms self-limiting micelles with exposed DNA-binding domains. We show that these micelles can stabilize DNA entanglements and in turn modulate DNA dynamics. Based on our findings, we conjecture that NANOG may contribute to regulate gene expression} by creating local gel-like environments that restrict genome dynamics and that its aging may ingrain mechanical memory in gene regulatory networks.  
}
\end{abstract}

\maketitle

\begin{figure*}[t]
	\begin{center}		\includegraphics[width=0.95\textwidth]{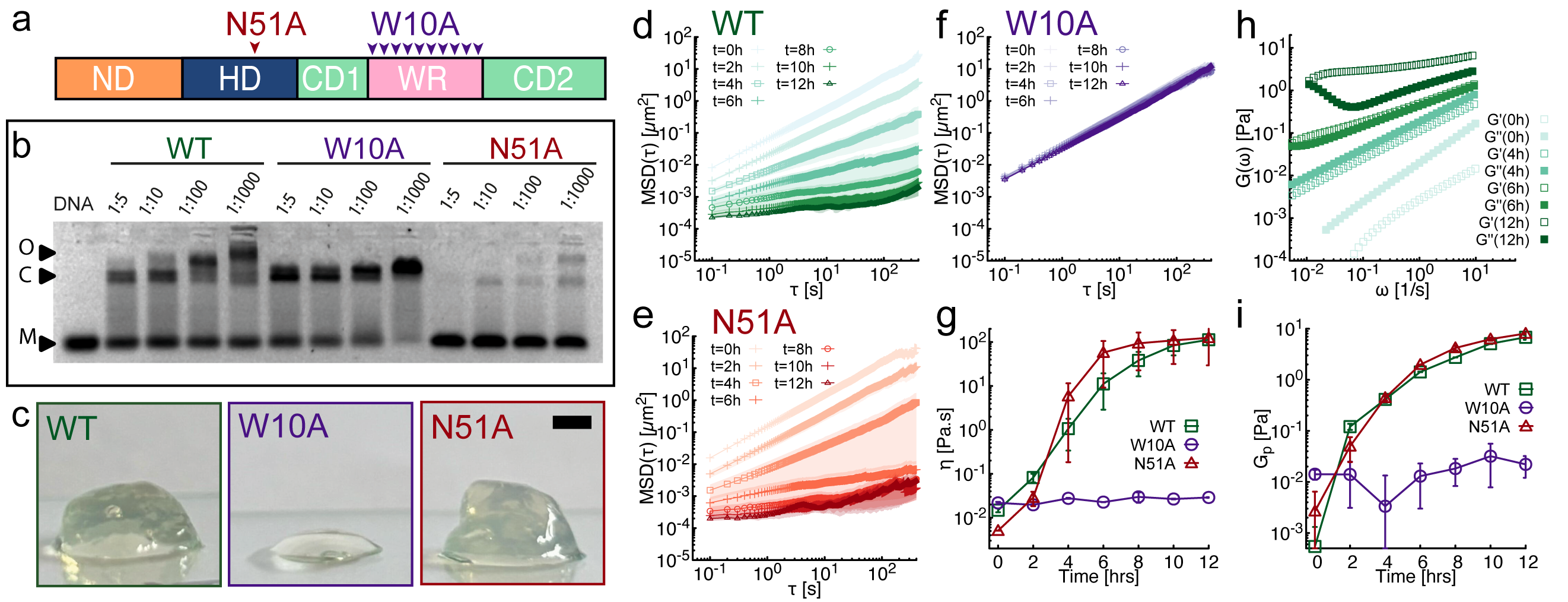}
    \vspace{-0.2 cm}
        \caption{\textbf{NANOG solutions form macroscopic aging gels.} \textbf{a.} NANOG protein is made of five domains: N-terminal domain (ND), DNA binding domain (HD), C-terminal domains (CD1,2) and an \dm{intrinsically disordered} tryptophan repeat (WR). Arrows point to the mutation sites for N51A and W10A. \textbf{b.} Electrophoretic mobility shift assay of NANOG mutants at different DNA:protein ratios. Each lane contains 5 nM of a 26-bp dsDNA oligomer. (M=monomer, C=DNA-NANOG complex, O=oligomer). \textbf{c.} Photographs of $\sim$mL samples of purified NANOG mutants aged at 1 mM overnight at 37$^\circ$C. WT and N51A form non-pipettable gels, whereas W10A remains liquid. \dm{These photos were taken after the samples were scooped out of Eppendorf tubes and could therefore be damaged}. Scale bar is 2 mm. \textbf{d-e-f.} Microrheology of the 3 NANOG mutants over the course of 12h at 1 mM and at 37$^\circ$C. \textbf{g.} Only WT and N51A show a $10^4$-fold increase in viscosity. \textbf{h.} Elastic ($G^\prime$) and viscous ($G^{\prime \prime}$) moduli for four aging times \dm{(see SI for more time points)}. \textbf{i.} WT and N51A fluids show significant increase in elastic plateau, computed as $G_p = G^\prime(\omega=10$ Hz).}
        \label{fig:aging_protein}
	\end{center}
    \vspace{-0.8cm}
\end{figure*}

Cell fate is regulated by transcription factors (TFs), proteins that bind DNA and regulate gene expression~\cite{alberts2022molecular}. NANOG is a TF required in embryonic stem cells (ESC)  to maintain a pluripotent state and prevent differentiation~\cite{Niwa2000,Chambers2003,Masui2007,Mitsui2003Nanog,KarwackiNeisius2013,Chambers2007}. However, the physical mechanisms through which NANOG controls cell fate in ESC are not known. 

\dm{Recent work has suggested that DNA-bridging~\cite{Choi2022}, liquid-liquid phase separation~\cite{Boija2018,McSwiggen2019,Gagliardi2013} and oligomerisation~\cite{Mullin2008,Wang2008} may underpin NANOG biological function. The former is enabled by NANOG's DNA binding domain, whereas the latter two require its intrinsically disordered region, or ``tryptophan repeat'' (WR, see Fig.~\ref{fig:aging_protein}a). Importantly, NANOG variants that display mutations in either the WR domain (W10A mutant) or the DNA binding homeodomain (N51A mutant) fail to maintain ESC pluripotency~\cite{Mullin2017,Novo2016Nanog} (see Fig.~\ref{fig:aging_protein}a). }

\dm{Currently, it is hypothesised that NANOG requires DNA binding and oligomerisation to bring together distant specific enhancers and promoters~\cite{Novo2018LongRange} to coordinate gene expression and maintain pluripotency. However, there is no evidence of major genome spatial re-organization following NANOG over-expression \textit{in vivo}~\cite{deWit2013Pluripotent,Choi2022}. Thus, it remains poorly understood why these domains are essential for ESC self-renewal. }

\dm{To better understand the physical principles underpinning NANOG function \textit{in vivo}, we decided to investigate the material and flow properties of NANOG and DNA-NANOG complex fluids. Specifically, we combined experiments and simulations to characterize the microscale assembly of mouse NANOG proteins\ and connected it to the macroscale properties of its biological complex fluids}. 

\dm{Our main discovery is that NANOG forms self-limited aging micelle-like amorphous clusters with surface-exposed DNA-binding domains, in turn acting as an aging cross-linker of DNA entanglements. In line with other intrinsically disordered DNA/RNA-binding proteins~\cite{Jawerth2020,Pulupa2025,Michieletto2022,Muzzopappa2021}, NANOG complex fluids display gelling and aging; however, opposite to previously seen unbounded condensates, NANOG gels are formed by self-limited micelles.} 

\dm{Our observations support the experimentally testable conjecture that NANOG micelle assemblies may be designed to regulate the genome's \textit{dynamics} (rather than its folding) by restricting the diffusion of key genomic sites and by ingraining mechanical memory through aging.}

\paragraph{Gelling and aging of NANOG solutions -- }
\dm{To better understand the physical principles underpinning the physiological functions of NANOG -- and inspired by other recent works on intrinsically disordered proteins~\cite{Jawerth2020,Michieletto2022} -- we investigated the flow properties of NANOG complex fluids.} We considered NANOG wild type (WT), its variant carrying a single-residue mutation in the DNA binding domain (N51A), and its mutant where all ten tryptophans in the WR are replaced by alanines (W10A). Both these mutants are not functional \emph{in vivo}, i.e. they lack the ability to maintain ESC self-renewal~\cite{Navarro2012NanogAutorepression,Mullin2017}, suggesting that these domains are essential to NANOG function.

To quantify DNA binding and oligomerization, we performed an electrophoretic mobility shift assay: NANOG proteins were incubated with a 26bp DNA fragment which is specifically bound by WT NANOG~\cite{Jauch2008} and then run on a gel. DNA bound to proteins migrate slower than free DNA, and we can thus visualize DNA binding and oligomerization of the mutants (Fig.~\ref{fig:aging_protein}b). As expected, the N51A mutant exhibits substantially reduced DNA-binding affinity (attenuation of DNA-protein complex (C) and oligomerisation (O) bands), while W10A lacks the oligomeric (O) band formed by the WT at high protein concentration~\cite{Mullin2017,Wang2008}. \dm{These results suggest that the intrinsically disordered WR domain is necessary for NANOG to create oligomeric complexes.}

\begin{figure*}[t!]
	\begin{center}
		\includegraphics[width=0.95\textwidth]{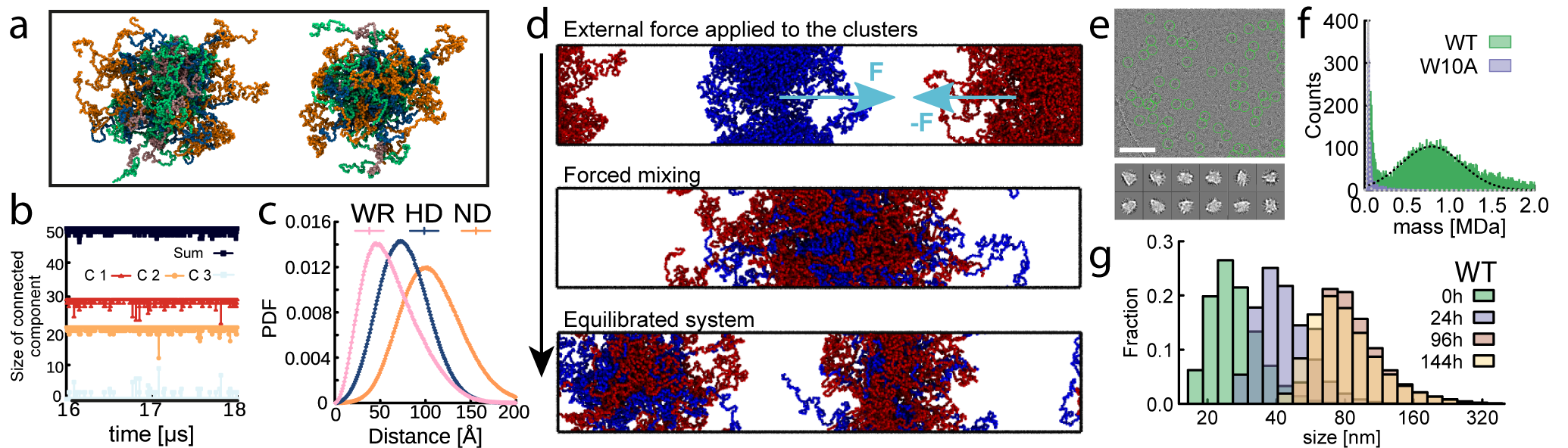}
        \vspace{-0.3cm}
		\caption{\textbf{NANOG forms micelle-like self-limited clusters through their WR}. \textbf{a.} Snapshot from equilibrated simulations of NANOG made of $M = 50$ proteins, each $N = 305$ amino acids long at $T = 300~\mathrm{K}$ and $\rho = 0.1~\mathrm{g/cm^3}$. Green, orange, pink and blue represent the different NANOG domains as in Fig.~\ref{fig:aging_protein}a. The system assembles into two disjoint clusters. \textbf{b.} Time evolution of the number of molecules forming the two main clusters and the third smaller cluster of molecules. \textbf{c.} Distribution of distances between the cluster COM to the residues in the different domains: N-terminal domain (ND, orange), homeodomain (HD, blue) and  tryptophan repeat (WR, pink). \textbf{d.} Simulated forced mixing of clusters (colored red and blue) and re-equilibration leading to the re-establishment of two disjoint clusters made of a different combination of molecules.  
        \textbf{e.} (Top) Representative cryo-EM motion corrected micrograph (scale bar 50 nm). (Bottom) 2D classes of objects showing heterogeneous and ``disordered'' assembly. Box size is 364 \AA, mask is 240 \AA, approximately the size of $\sim$20-30 NANOG molecules. \textbf{f.} Mass photometry showing peaks at weight corresponding to $\sim$22-25 NANOGs per cluster. \textbf{g.} DLS quantifying the hydrodynamic size of clusters in samples of NANOG incubated at 37$^\circ$C and different times.
        }
        \label{fig2}
	\end{center}
    \vspace{-0.9cm}
\end{figure*}

\dm{Given the analogy between the WR domain of NANOG and that of other proteins with intrinsically disordered domains, we decided to investigate if the WR was required for NANOG to display non-Newtonian liquid properties, such as gelling, phase separation or aging~\cite{Jawerth2020,Michieletto2022}}. We thus incubated 1mM NANOG WT, N51A and W10A at 37$^\circ$C overnight and noticed that both WT and N51A underwent a sol-gel transition while W10A remained a pipettable liquid (Figure~\ref{fig:aging_protein}c). 
\dm{Since performing a systematic bulk rheology study on $\sim$mL of NANOG protein at high concentration is not feasible, we turned to microrheology~\cite{Mason1995,Mason1997ParticleTrackingMicrorheology,Mason2000} to quantify the gelation timescales and the gels' rheological properties}. \dm{Microrheology is well suited to sample the passive rheological behaviour of small quantities ($\sim \mu$L) of fluids through short ($\sim 2$ min) videos, which are much shorter than the typical aging time of protein fluids~\cite{Jawerth2020}. On the contrary, bulk rheology and active microrheology would require larger sample volumes or longer measurement times~\cite{Michieletto2022}}. Using microrheology, we measured the 2D mean squared displacement (MSD) of passive tracers embedded in the NANOG solutions as $\langle \Delta r^2 (\tau) \rangle = \langle \left[\bm{r}(t_0+\tau) - \bm{r}(t_0)\right]^2 \rangle$, where $\bm{r}(t)$ is the position of the tracer at time $t$, and the average is performed over tracers and initial times $t_0$. The tracers in both 1mM WT and N51A displayed subdiffusive behavior ($\langle \Delta r^2 \rangle \sim t^\alpha$ with $\alpha<1$) within 4-6 hours of incubation at 37$^\circ$C (Fig.~\ref{fig:aging_protein}d,e), in turn indicating the onset of elasticity and gelling~\cite{Michieletto2022,Cicuta2007}. On the contrary, tracers embedded in the 1mM W10A solution remained freely diffusive ($\langle \Delta r^2 \rangle \sim t$) at all times (Fig.~\ref{fig:aging_protein}f). This strongly suggests that loss of tryptophans in the WR, which abolish oligomerisation~\cite{Mullin2017}, also prevent gel formation. After 12h of incubation the effective viscosity of the WT and N51A solutions \dm{-- obtained as $\eta = k_BT/(3 \pi a D)$ with $a$ the diameter of the tracer and $D = \lim_{t \to \infty} \langle \Delta r^2 (t)\rangle/4t$ the diffusion coefficient~\footnote{We tracked trajectories independently in the x and y directions and averaged them, so that the MSD is expected to be proportional to $2Dt$ for 1D free diffusion. Where the MSD was not scaling linearly with time, we approximated the diffusion coefficient with its upper bound based on the measured MSD.} --} was $10,000$-fold larger than the original solution (Fig.~\ref{fig:aging_protein}g). \dm{This result was unaffected by the tracers' size and passivation considered (see SI~\cite{SI}).}

We further \dm{quantified} the aging process through the generalized Stokes-Einstein relation~\cite{Mason2000,Mason1995,Cicuta2007,Brizioli2025OFM}, $G^*(\omega) = (k_B T)/(\pi a i \omega \mathcal{F}_u \left[ \langle \Delta r^2 \rangle \right])$, which allows us to compute the elastic and viscous shear moduli ($G^\prime(\omega)$ and $G^{\prime \prime}(\omega)$) from the MSDs. In Fig.~\ref{fig:aging_protein}h, we show that around 5 hours, the elastic modulus dominates over the viscous one across frequencies, in turn identifying this as the gelation point. The elasticity of the WT and N51A samples also grows $10,000$-fold over 12 hours, however the W10A mutant displays no onset of elasticity (Fig.~\ref{fig:aging_protein}i), in line with the previous observation that WR is necessary to mediate NANOG oligomeric assembly.

\dm{Interestingly, we do not observe sol-gel transition at room temperature for 1mM WT protein solution (see SI, Fig.~S5). This suggests that the aging process is facilitated by the re-configuration of the intrinsically disordered WR domain within NANOG oligomeric assemblies, which can more readily explore a rugged free energy landscape at higher temperatures. This behaviour is in line with the aging Maxwell fluids model~\cite{Jawerth2020}; however, our time-cure superposition analysis does not align with the one expected for simple Maxwell fluids and instead displays multiple characteristic timescales which we will explore further in the future (see SI Fig.~S2, for details)}. 

\paragraph{Simulations reveal NANOG assembles into self-limiting micelles -- } To better understand the gelation observed in Fig.~\ref{fig:aging_protein}, we decided to simulate NANOG proteins using the Mpipi framework: a residue-level model parameterized on both bioinformatics data and all-atom simulations~\cite{Mpipi2021}. We simulated $M = 50$ NANOG proteins, each consisting of $N = 305$ amino acids. The proteins were initially uniformly distributed in an elongated simulation box corresponding to an overall density of $\rho = 0.1~\mathrm{g/cm^3}$. The system was then evolved in the NVT ensemble using LAMMPS~\cite{LAMMPS2022} with implicit solvent at a temperature of $T = 300$ K. A time step of $10~\mathrm{fs}$ was used for integration over approximately $2 \times 10^9$ steps (i.e., $20~\mu$s) and we ran 3 independent replicas (see SI for details). We observed that at equilibrium, NANOG WT proteins assembled into clusters characterized by a well-defined size of approximately 30 proteins (see Fig.~\ref{fig2}a). \dm{On the contrary, simulations of W10A displayed no clusters (see SI, Fig.~S1)}. Although the clusters' composition was dynamic (Fig.~\ref{fig2}b), the overall number of molecules per cluster remained the same through the simulation, in turn suggesting that the proteins assembled in self-limited, yet dynamic, clusters. To further characterize the internal structure of NANOG assemblies, we computed the distances of the domains from the center of mass of the cluster. The radial distributions reveal that NANOG molecules form a micelle-like structure, where the WR occupies the cluster core, the DNA binding domain occupies the intermediate layer and the N-terminal domain forms the outer layer (Fig.~\ref{fig2}c). Similar micelle-like organization was observed in simulations of human NANOG, however the observed cluster size was larger~\cite{Takada2023}, \dm{possibly because of the different number of tryptophans in the domain.}

To test the stability of the assemblies we artificially forced mixing of the clusters into one larger cluster, removed the mixing forces, and let the system re-equilibrate for $20\,\mu\text{s}$ (Fig.~\ref{fig2}d). Strikingly, we observed the re-organization of the assemblies into two distinct clusters formed by two new subsets of proteins. This clearly suggests that NANOG may behave similarly to a block co-polymer, forming thermodynamically stable self-limited micelle-like assemblies~\cite{Hamley2007}.

\paragraph{Experimental validation of self-limiting NANOG assemblies -- }
\dm{To experimentally test the prediction of our simulations we performed cryo-EM, mass photometry and dynamic light scattering}. \dm{Cryo-EM reproducibly identified NANOG complexes that were monodisperse in size, approximately 25 nm and compatible with the size expected for a globular assembly of 20-30 NANOG monomers (Fig.~\ref{fig2}e). However, the assemblies were heterogeneous in shape and did not lock into a discrete conformational space; in turn, we could not solve their structure~\cite{relion}. This suggests that NANOG complexes assume an amorphous internal organisation, which is typical of intrinsically disordered proteins \dm{and is in agreement with our simulations}. Mass photometry also identified assemblies with molecular weight $M_W \simeq 0.8$ MDa, compatible with $\simeq$ 22-25 NANOG WT monomers (Fig.~\ref{fig2}f)}. \dm{Finally, we performed dynamic light scattering at different aging times, during 37$^\circ$C incubation of 30 $\mu$M NANOG sample. The distributions in Fig.~\ref{fig2}g display a clear shift towards larger hydrodynamic size with time, however the shift stops after 96h incubation, confirming the self-limiting nature of the NANOG clusters. This data also supports the idea that the proteins do not aggregate due to unfolding at these temperatures, as otherwise the clusters would keep growing indefinitely.}

\begin{figure}[t!]
	\begin{center}
\includegraphics[width=0.45\textwidth]{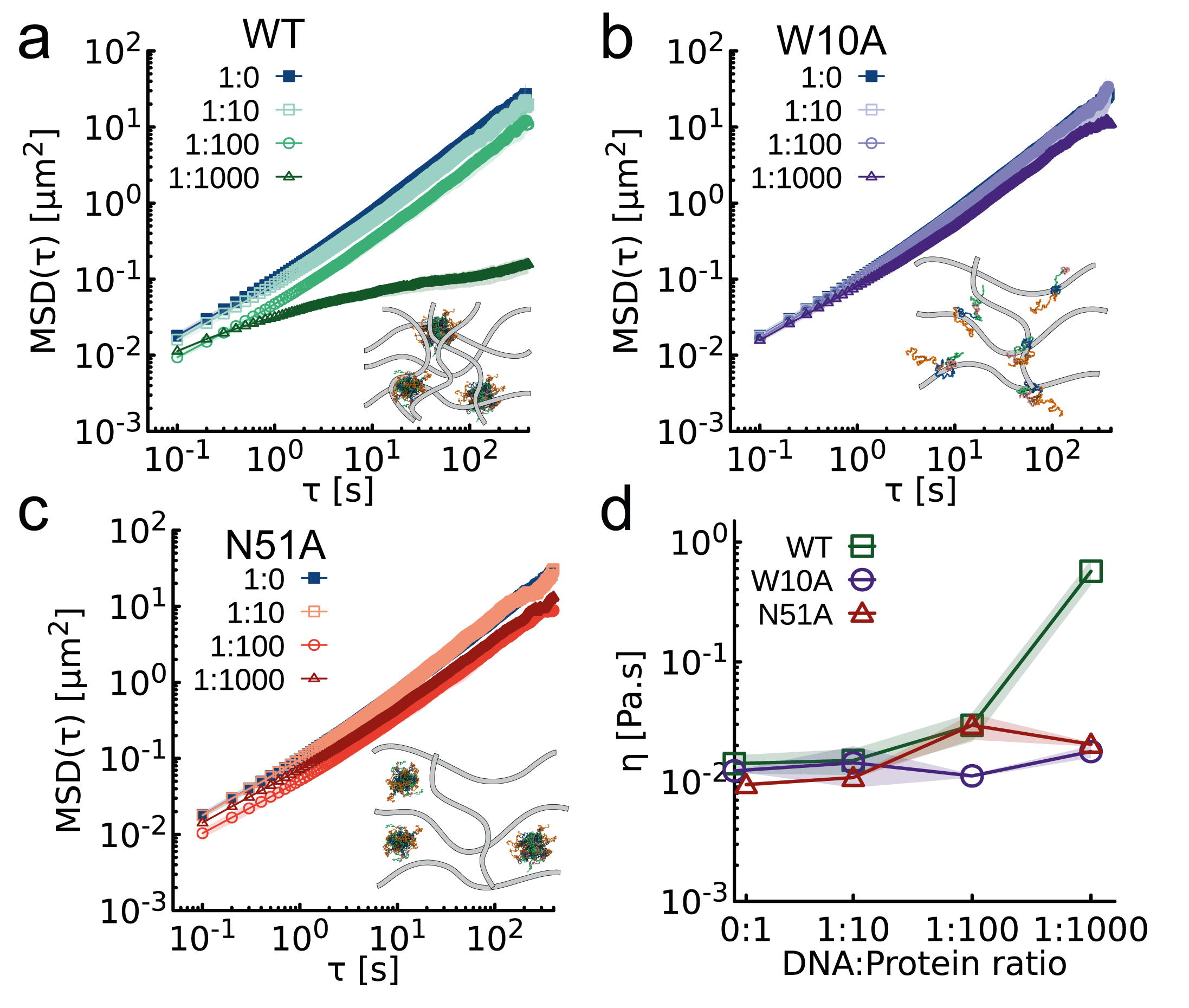}
        \vspace{-0.4cm}
		\caption{\textbf{NANOG micelles can crosslink DNA}. 
        \textbf{a-c.} MSD of tracers at varying DNA:protein stoichiometries in a solution of \ldna at 7.9~nM and (a) WT, (b) W10A and (c) N51A, at different concentrations. Insets depict schematics of NANOG assemblies in the presence of \ldna. \textbf{d.} Viscosity of DNA–protein solutions as a function of protein concentration. \dm{Measurements are performed at room temperature and immediately after mixing.}}
        \label{fig3}
	\end{center}
    \vspace{-1 cm}
\end{figure}

\dm{Based on this evidence, we now interpret the sol-gel transition observed in Fig~\ref{fig:aging_protein} as due to the formation of sticky self-limited micelles which overlap at high concentrations and form a percolating network that drives gelation (consistent with our simulations, see Fig.~S1c). In support of this interpretation, and based on the measured hydrodynamic size of the micelles ($\sigma \simeq 80$ nm) and their molecular weight ($M_W \simeq 0.8$ MDa), we estimate the overlapping concentration as $c^* = 3 M_W/(4 \pi N_A m_W (\sigma/2)^3) \simeq 0.14$ mM ($m_W = 34$ kDa is the molecular weight of a NANOG monomer), smaller than the concentration of 1 mM at which gelling is observed in Fig.~\ref{fig:aging_protein}.}


\paragraph{NANOG-DNA solutions form aging gels at physiological concentrations -- } To better understand the physiological role of the observed gelling, we performed microrheology~\cite{Mason1995,Zhu2008} on solutions of \ldna (48,502 bp) at $c = 250$ ng/$\mu$l (or 7.9 nM) mixed with NANOG at physiological concentrations. \dm{We considered NANOG concentrations between 10nM and 10 $\mu$M, in broad agreement with the values reported \emph{in vivo}, 80-750 nM ~\cite{Choi2022}. However, we also know that NANOG is heterogeneously distributed in cells and forms phase-separated foci with higher local concentration, likely reaching $\simeq$1-10 $\mu$M~\cite{Boija2018}. Finally, the DNA concentration considered in our experiments is around 10-fold larger than \ldna overlapping concentration ($c^* \simeq 20$ ng/$\mu$l~\cite{Fosado2023ihf}) and the DNA is therefore entangled, mimicking the crowded nuclear environment. }

The tracers embedded in solutions of DNA and WT NANOG at room temperature and without pre-incubation display a marked, and abrupt, sub-diffusive behaviour at around 7.9 $\mu$M NANOG concentration (Figs.~\ref{fig3}a-c). In contrast, solutions made with W10A and with N51A mutants did not display a comparable change in MSDs over the same concentration range.
We rationalize this results as follows: NANOG WT forms micelles where the DNA-binding domains is exposed, enabling each micelle to bridge multiple \ldna molecules and in turn creating an elastic gel (Fig.~\ref{fig3}a). The mutants either cannot form micelles (W10A), or do not bind DNA as strongly (N51A), and therefore they cannot crosslink DNA as effectively (Fig.~\ref{fig3}b-c). This behaviour is akin to that of other protein condensates which show solid-like behaviour when mixed with DNA/RNA~\cite{Muzzopappa2021,Michieletto2022,Shin2017}, \dm{however contrary to forming unbounded condensates, NANOG forms self-limited micelles}.

\begin{figure}[t!]
	\begin{center} \includegraphics[width=0.45\textwidth]{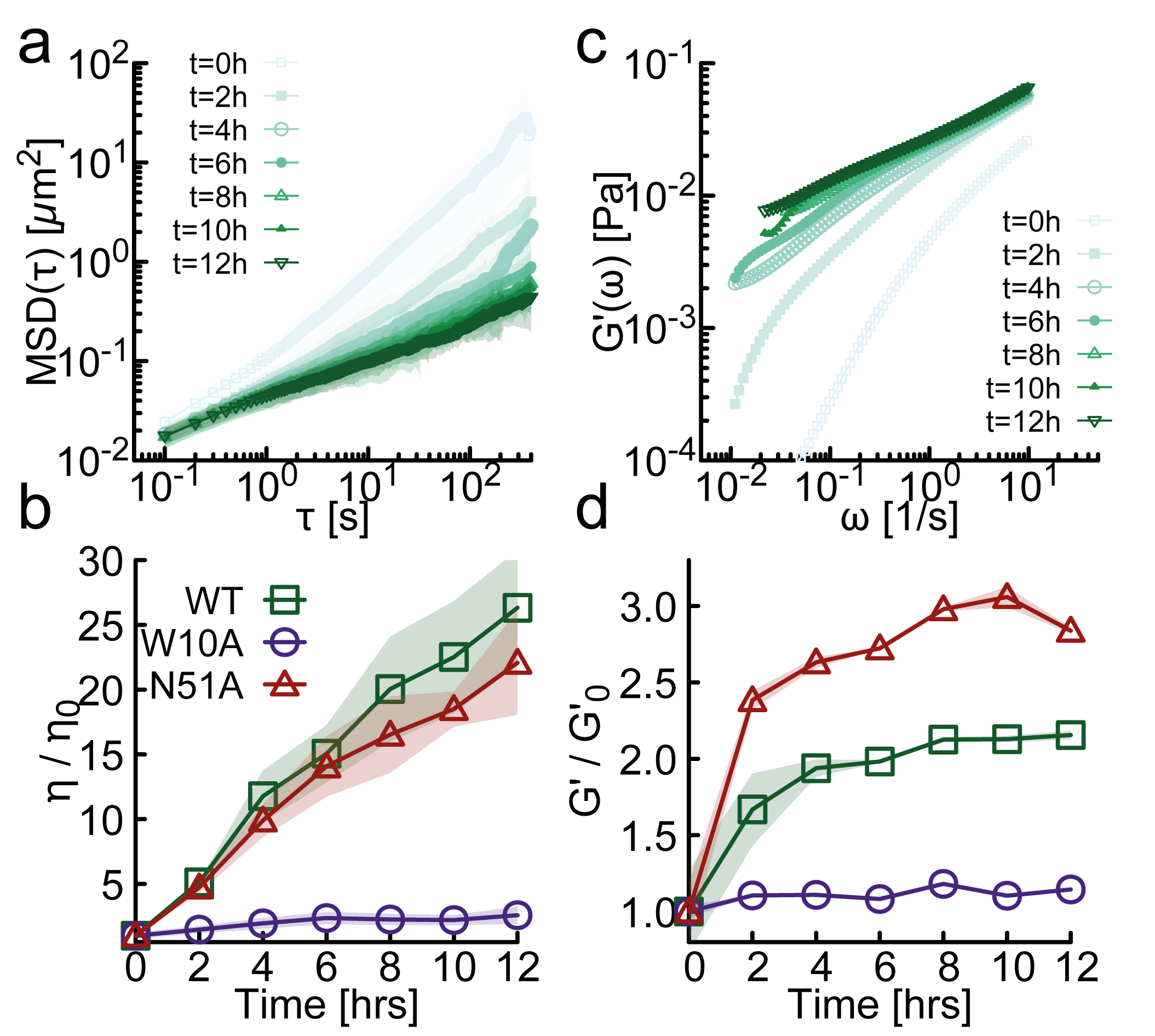}
        \vspace{-0.3cm}
		\caption{\textbf{NANOG-DNA solutions gel at physiological concentration and temperature}. 
        \textbf{a.} Time resolved MSDs of tracers embedded in solutions of NANOG (7.9~$\mu$M) and \ldna (7.9 nM). \textbf{b.} Normalized viscosity of \ldna-NANOG solutions obtained with different mutants (normalized to the initial viscosity $\eta_0$). \textbf{c.} Time resolved elastic modulus $G^\prime$. \textbf{d.} Elastic modulus plateau $G_p = G^\prime (\omega=10 \textrm{Hz})$ as a function of aging time (normalized to the initial elastic plateau modulus $G^\prime_{0}$). \dm{Measurements are performed at 37$^{\circ}$C.}} 
        \label{fig4}
	\end{center}
    \vspace{-0.8cm}
\end{figure}

\dm{Finally, we asked if the solutions of entangled \ldna and NANOG could display aging at physiological concentrations and temperatures. We thus performed microrheology on solutions of \ldna (7.9nM) mixed with NANOG mutants (7.9 $\mu$M) at 37$^{\circ}$C (see Fig.~\ref{fig4} and SI Fig.~S3). Interestingly, the presence of DNA triggers aging and gelation despite using an 18-fold lower protein concentration than the estimated overlap concentration (0.14 mM) and a 100-fold lower concentration than in Fig.~\ref{fig:aging_protein}}. \dm{This result suggests that \textit{in vivo}, NANOG foci may act as solid-like gel microenvironments that restrict DNA dynamics. From a polymer physics perspective, this is an unconventional gelation process because it is driven by micelles that act as aging crosslinkers and whose strength increases over time.}

\paragraph*{Conclusions --} We combined experiments and simulations to understand the biophysical properties of a key transcription factor, NANOG, involved in maintenance of pluripotency in ESC. 

\dm{First, we discovered that at concentrations above its overlapping concentration ($c^*\simeq 0.14$ mM) NANOG WT forms an aging macroscopic gel (Fig.~\ref{fig:aging_protein}). Residue-level Molecular Dynamics simulations revealed that NANOG form self-limited micelle assemblies with amorphous internal structure and exposed DNA-binding domains (Fig.~\ref{fig2}a-d). Additionally, they suggest that the intrinsically disordered domain is required to form these micelles (see SI Fig.~S1). We confirmed this prediction experimentally (Fig.~\ref{fig2}e-g): NANOG WT assemble in self-limited amorphous clusters of 22-25 monomers.
On the other hand, W10A does not form micelles. We therefore interpret the gelation as due to these disordered micelles overlapping with each other and forming a percolating network which drive a macroscopic sol-gel transition.}

\dm{To understand how NANOG modulates genome organization \textit{in vivo}, we performed microrheology experiments on mixtures of NANOG and $\lambda$DNA near physiological conditions (Fig.~\ref{fig3}-\ref{fig4}). We discovered that in the presence of DNA, NANOG WT can form gels at physiological concentrations (1-10 $\mu$M) and temperatures (37$^\circ$C). We conjecture that this \dm{(DNA length-dependent, see SI Fig.~S4)} phenomenon is driven by the NANOG micelles acting as multivalent crosslinkers, in turn stabilising entanglements and restricting DNA dynamics. The NANOG-DNA solutions also displayed aging, which we attribute to the molecular reorganization of the intrinsically disordered domains within the assembly, exploring a rugged free energy landscape.}


We argue that the gelling behavior displayed by NANOG-DNA solutions may be functionally relevant: it allows NANOG to restrict the dynamics of key genomic sites and to stabilise a gene regulatory network. In this model (which could also explain similar sol-gel transitions seen in other TFs~\cite{Pulupa2025}) NANOG would modulate genome \textit{dynamics}, rather than its 3D organization. Further, we conjecture that the aging of NANOG micelles may physically ingrain memory in a regulatory gene network by physically reinforcing genomic interactions. 


\dm{To test our predictions \textit{in vivo}, it would be interesting to simultaneously track fluorescently labeled NANOG and chromatin loci. Our hypothesis is that areas where NANOG is enriched (foci) should correlate with slow chromatin dynamics. If proved correct, our model would reveal a new paradigm through which some TFs may regulate gene expression by modulating the dynamics of the genome, rather than its folding.} In light of this model, it will also be interesting to investigate the combined effect of NANOG with different TFs, such as Sox2, Oct4, and Klf4~\cite{Takada2025}, other gel-forming RNA-binding nuclear proteins~\cite{Michieletto2022,Marenda2024} and their influence on chromatin structure and dynamics~\cite{Li2025Chromatin}.

\section{Acknowledgements}
YAGF acknowledges support from the Physics of Life, UKRI/Wellcome (grant number EP/T022000/1–PoLNET3). DM acknowledges the Royal Society and the European Research Council (grant agreement No 947918, TAP) for funding. IC acknowledges grant support from the MRC (MR/T003162/1) and BBSRC (BB/T008644/1). We also acknowledge support from MRC Precision Medicine PhD programme and Wellcome Trust integrated cellular mechanisms PhD programme. 
We acknowledge the Scottish Centre for Macromolecular Imaging, Mhairi Clarke and James Streetley for assistance with cryo-EM experiments and access to instrumentation, financed by the Medical Research Council (MC-PC-17135) and the Scottish Funding Council (H17007). For the purpose of open access, we have applied a Creative Commons Attribution (CCBY) licence to any author accepted manuscript version arising from this submission.

\bibliography{bibliography}
\end{document}